\documentclass[twocolumn,aps,prb,groupedaddress,showpacs]{revtex4}
\usepackage{graphicx}

\newcommand{\y}{\kappa}

\begin{document}

\title{Two dimensional electron liquid 
in the presence of Rashba spin-orbit coupling: 
symmetric momentum space occupation states}
\author{Stefano Chesi}
\author{George Simion}
\author{Gabriele F. Giuliani}
\affiliation{Department of Physics, Purdue University,
West Lafayette, IN 47907, USA}

\date{\today}

\begin{abstract}
The orientation of the local electron spin quantization axis in momentum space 
is identified as the most significant physical variable in determining the
states of a two-dimensional electron liquid in the presence of Rashba spin-orbit 
coupling. Within mean field theory several phases can be identified that are
characterized by a simple symmetric momentum space occupation. The problem admits 
uniform paramagnetic as well as spin polarized chiral solutions. The latter have 
a nontrivial spin texture in momentum space and are constructed out of states 
that are not solutions of the non interacting Hamiltonian. The concept of generalized
chirality as well as the stability of spatially homogeneous states are also discussed.
\end{abstract}


\maketitle
Recent interest in the physics of two dimensional electronic systems subject to the
effects of Rashba spin-orbit interaction makes the corresponding many-body problem one
of timely interest.
Since it is well known that interaction effects in modern quasi-two dimensional 
electron liquid devices are quite strong in a wide range of 
accessible densities,\cite{tan05,winkler05,asgari05,EP2DS15_05_proc} 
it is clearly of fundamental importance to establish a sound set of theoretical 
notions about the effects of the electron-electron interactions in this problem. 
Patterning the general approach on that followed for the familiar case of the 
electron liquid in the absence of spin-orbit,\cite{TheBook} the first step 
in tackling the many-electron problem is to establish a meaningful mean field theory.
Once this step is successfully taken, and the relevant dynamical variables and 
symmetries identified, correlation corrections can be eventually approximately calculated 
by a variety of methods. The purpose of this paper is to describe some of the general 
results of the mean field theory of the two dimensional electron liquid in the presence 
of Rashba spin-orbit.\cite{comment_generalized_SO}
The present analysis is limited to the simple case of symmetric momentum space 
occupation states that, as we will show, can be completely and elegantly 
classified in terms of what we will refer to as the generalized chirality.

\begin{figure}
\includegraphics[width=0.35\textwidth]{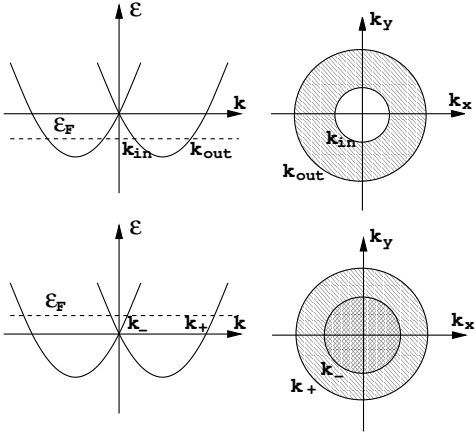}
\caption{\label{occupation} Two distinct, symmetric compact ways of occupying non 
interacting chiral states in $\mathbf{k}$ space. In the top panel only the lowest
chiral band is occupied and the generalized chirality differs from the ordinary
chirality which in this case is unity. The left panel represents the non interacting 
energy spectrum.}
\end{figure}

The problem is defined by the following model hamiltonian:
\begin{equation}
\label{Hint}
\hat H ~=~ \sum_i \hat H_0^{(i)}+\frac{1}{2}\sum_{i\neq j}
\frac{e^2}{| \hat{\bf r}_i- \hat{\bf r}_j|} ~,
\end{equation}
where the electronic motion is limited to the $x-y$ plane and the single particle 
terms contain a spin-orbit interaction of the Rashba type\cite{rashbaSO,comment_dresselhaus}
\begin{equation}
\label{H0}
\hat{H}_0 ~=~ 
\frac{ \hat{{\bf p}}^2}{2m} 
+ \alpha~(\hat{\sigma}_x \hat{p}_y - \hat{\sigma}_y \hat{p}_x) ~,
\end{equation}
where $\alpha$ is assumed positive and terms corresponding to an homogeneous 
neutralizing background are understood. 
In the homogeneous case $\hat{H}_0$ has eigenfunctions and eigenvalues given by
\begin{equation}
\label{phi0kpm}
\varphi_{ {\bf k} , \pm} ( {\bf r} ) =
\frac{e^{i {\bf k} \cdot {\bf r} }}{\sqrt{2 L^2}}
\left(
\begin{array}{c}
\pm 1\\
i e^{i\phi_{\bf k}}
\end{array}
\right)~,\qquad 
\epsilon_{\mathbf{k}\pm } = 
\frac{\hbar^2 \mathbf{k}^2}{2 m}\mp \alpha \hbar \, k ~,
\end{equation}
where $L$ is the linear size of the system and $\phi_{\bf k}$ is the angle 
between the direction of the wave vector and the $x$-axis. Interestingly 
spin-orbit forces each state in momentum space to have its own spin quantization axis. 
This direction lies in the $x-y$ plane and makes an
angle of $\frac{\pi}{2}$ with ${\bf k}$. The corresponding unit vector will be 
denoted by $\hat{\phi}_\mathbf{k}$. 
These states form two split bands characterized by opposite chirality 
which are schematized in Fig.~\ref{occupation}.\cite{rashbaSO} 

Since the mean field theory involves, but as we shall see is not limited to, 
single Slater determinants obtained by occupying the single particle states 
(\ref{phi0kpm}), it proves necessary to establish a complete yet manageable 
classification scheme.
We have found that for states with isotropic, compact occupation in momentum 
space the following quantity, 
\begin{equation}
\label{gen_chirality_def}
\chi=
\left\{
\begin{array}{cl}
\chi_0 \qquad &   \mathrm{for\,\, } 0 \leq \chi_0  < 1  \\
\frac{k_{out}^2+k_{in}^2}{k_{out}^2-k_{in}^2} \qquad & \mathrm{for \,\,} \chi_0= 1 
\end{array}
\right.  ~,
\end{equation}
which we will refer to as the {\it generalized chirality}, offers such possibility.
In (\ref{gen_chirality_def}) $\chi_0=\frac{N_ + -N_-}{N_+ + N_-}$ is the 
ordinary chirality as defined in terms of the occupation $N_\pm$ of each band, 
while $k_{in}$, $k_{out}$ and $k_{\pm}$ are geometrical parameters in momentum 
space characterizing the occupied regions (circles for the isotropic case) and are 
defined in Fig.~\ref{occupation}.
Depending on the relative occupation of the bands, $\chi$ can acquire any value 
from zero to infinity and allows one to uniquely label the relevant 
set of many-particle states for fixed density $n= N/L^2$. 
This must be contrasted with the ordinary chirality which satisfies the condition 
$ \chi_0 \leq 1$ and equals unity for all the states in which only one of the 
chiral bands is occupied. 

\begin{figure}
\includegraphics[width=0.2\textwidth]{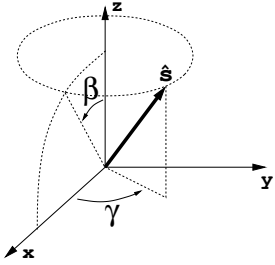}
\caption{\label{essedikappa} Geometry of the local (in momentum space) spin 
quantization axis. The $z$-axis is taken perpendicular to the plane of motion.}
\end{figure}

Limiting for the time being our analysis to spatially homogeneous solutions, it is
readily established that, because of the peculiar physics of the Rashba spin-orbit, 
the momentum space local orientation of the electron spin quantization axis, 
as determined by its unit vector $\hat{s}_{\mathbf{k}}$, plays a crucial physical 
role. In Fig.~\ref{essedikappa} the angles $\beta_{\mathbf{k}}$ and $\gamma_{\mathbf{k}}$ 
determining the orientation of $\hat{s}_{\mathbf{k}}$ are defined.
Accordingly our idea is to construct many-particle Fock states 
$\Psi [n_{\mathbf{k}\,\pm }, \hat{s}_{\mathbf{k}}]$ with occupations $n_{\mathbf{k}\,\pm }$ 
using single particle plane wave states each characterized by its own $\hat{s}_{\mathbf{k}}$. 
Our main task is therefore reduced to determine what, for a given density, is the 
most energetically favorable set of $n_{\mathbf{k}\,\pm }$ and corresponding spin 
quantization axes $\hat{s}_{\mathbf{k}}$.

From a formal point of view this problem can be handled by making use of a 
standard Wick decoupling procedure.\cite{TheBook} We find however more instructive 
and elegant to minimize the total energy as a functional of both 
$n_{\mathbf{k}\,\pm }$ and $\hat{s}_{\mathbf{k}}$. 
It is a simple exercise to write down the expectation 
value of the full hamiltonian of the electron liquid (\ref{Hint}) taken over a 
generic single Slater determinant 
$\Psi [n_{\mathbf{k}\,\pm }, \hat{s}_{\mathbf{k}}]$: 
\begin{eqnarray}
\label{HFenergy}
E_\Psi [n_{\mathbf{k}\,\mu} , \hat{s}_{\mathbf{k}} ]
=\sum_{\mathbf{k};\,\mu = \pm} \left(
\frac{\hbar^2 \mathbf{k}^2}{2m}\,n_{\mathbf{k}\,\mu } -
\hbar \alpha 
\mu \, k \, \hat{\phi}_\mathbf{k}\cdot\hat{s}_{\mathbf{k}}\, n_{\mathbf{k}\,\mu}
\right) \nonumber \\
-\frac{1}{4 L^2}\sum_{\mathbf{k},\mathbf{k}';\,\mu ,\mu' = \pm} 
v_{\mathbf{k}-\mathbf{k}'}\,
( 1+ \mu \mu'\,\, \hat{s}_{\mathbf{k}}\cdot \hat{s}_{\mathbf{k}'}) \,
n_{\mathbf{k}\,\mu}n_{\mathbf{k}'\mu'} ~.
\end{eqnarray}
The corresponding, renormalized, single particle energies 
$\epsilon_{\mathbf{k}\,\mu}$
can then be readily obtained by differentiation with respect to 
$n_{\mathbf{k}\,\pm }$.\cite{comment_previous_work_QP}

\begin{figure}
\begin{center}
\includegraphics[width=0.4\textwidth]{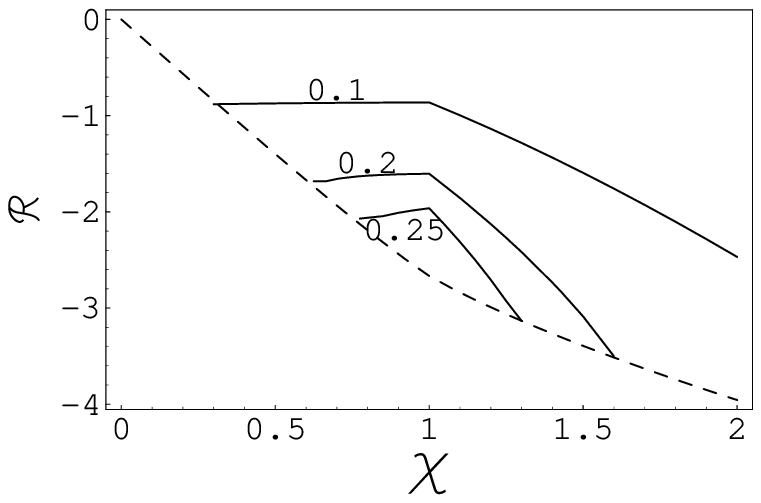}
\includegraphics[width=0.4\textwidth]{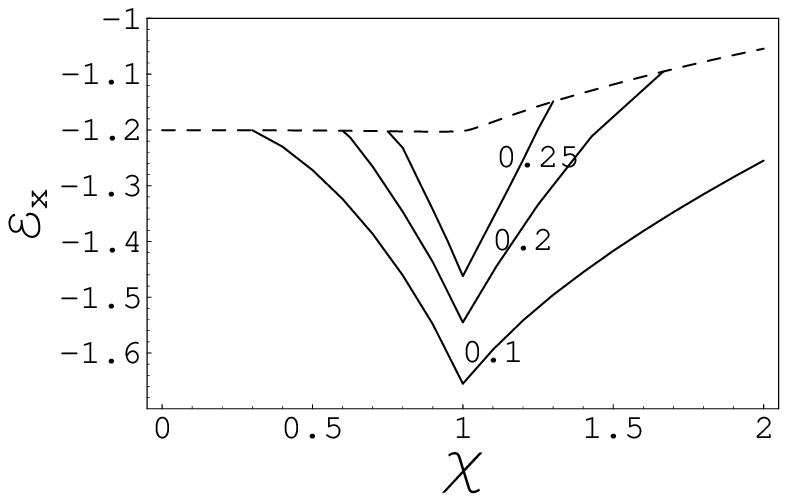}
\caption{
\label{RandEx}
Plot of $\mathcal{R}[\chi; \bar \beta]$ (top) and
$\mathcal{E}_{x}[\chi;\bar \beta]$ (bottom), as defined in (\ref{Egen}), evaluated
for the mean field solutions determined by solving Eq.~(\ref{beta_eqn_adim}) as
functions of the generalized chirality $\chi$ for different values of $\bar\alpha$.
Dashed lines: unpolarized case. Solid lines: transverse ferromagnetic solutions.
}
\end{center}
\end{figure}

We begin our analysis of Eq.~(\ref{HFenergy}) by noting that, as one can show, for 
symmetric occupations, the case $\gamma_{\bf k}=\phi_{\bf k}+\frac{\pi}{2}$ corresponds 
to an energy minimum. In this case by implementing the momentum rescaling
\begin{equation}
\label{scaled_beta}
\beta_{\bf k}=\bar\beta(\frac{|{\bf k}|}{\sqrt{2\pi n}}) ~,
\end{equation}
the functional $E_\Psi$ can be expressed in Rydberg units as follows 
\begin{equation}
\label{Egen}
{\cal E}_\Psi [n_{\mathbf{k}\,\mu} , \hat{s}_{\mathbf{k}} ] =
\frac{\mathcal{K}(\chi)}{r_s^2}
+ \bar{\alpha} \frac{\mathcal{R}[\chi; \bar \beta]}{r_s} +
\frac{\mathcal{E}_{x}[\chi;\bar \beta ]}{r_s} ~,
\end{equation}
where we have also introduced the dimensionless parameters 
$r_s^{-1} = \sqrt{\pi a_B n}$ and $\bar\alpha=\frac{\hbar\alpha}{e^2}$. 
This formula is quite remarkable for it displays a simple explicit dependence 
on the spin-orbit coupling constant and an high degree of universality. Specifically 
$\mathcal{K}$ is a simple universal function of $\chi$: 
\begin{equation}
\label{KE}
\mathcal{K}(\chi)=
\theta (1- \chi) (1+\chi^2) ~+~  \theta (\chi - 1) 2 \chi ~,
\end{equation}
while $\mathcal{R}$ and $\mathcal{E}_{x}$ are in addition universal 
functionals of the rescaled momentum dependent angle $\bar \beta$ 
whose expressions are reported elsewhere.\cite{unpub_ChesiGFG}
For paramagnetic states the situation simplifies further since for this class 
of solutions $\beta_\mathbf{k} = \bar \beta = \frac{\pi}{2}$ and only the dependence on
$\chi$ survives. Figure \ref{RandEx} displays the typical dependence of 
$\mathcal{R}$ and $\mathcal{E}_{x}$ on $\chi$ for actual solutions 
$\bar \beta$ of the problem for different values of $\bar\alpha$. 
As we show below, for a given value of the latter, the final values of 
$\mathcal{R}$ and $\mathcal{E}_{x}$ can be expressed as a function of $\chi$ only.
Notice the cuspy behavior about the critical value $\chi = 1$ which separates states
for which one or two (renormalized) chiral bands are occupied.

\begin{figure}
\begin{center}
\includegraphics[width=0.4\textwidth]{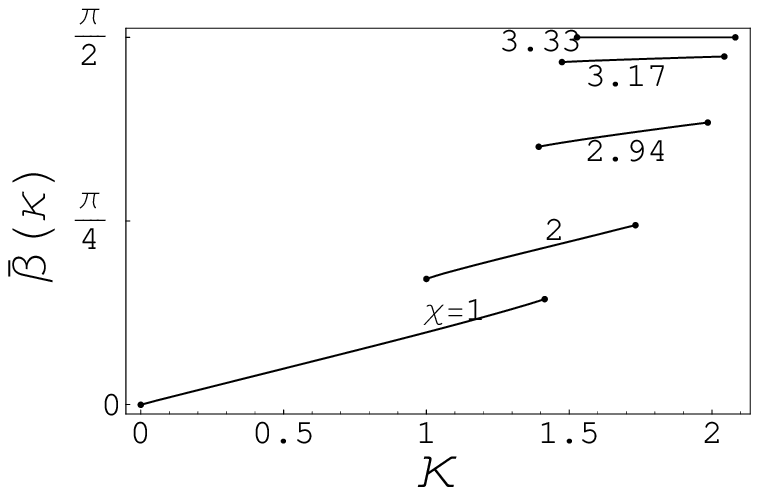}
\includegraphics[width=0.4\textwidth]{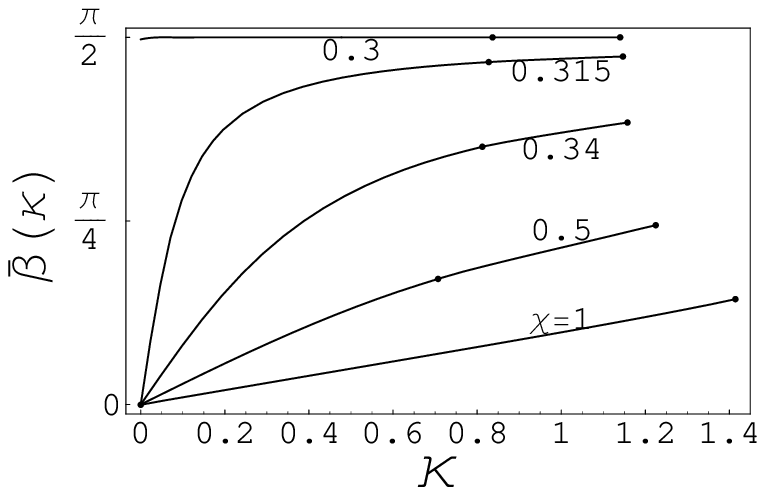}
\caption{\label{manybeta} Top: universal spin azimuthal 
angle $\bar\beta(\y)$ for the case of $\bar{\alpha}=0.1$ 
and various values of $\chi$. 
Bottom: same for the corresponding reciprocal values of $\chi$. 
The dots mark the ending points of the $[\sqrt{|1-\chi|},\sqrt{1+\chi}]$ intervals
corresponding to the occupied regions of Fig.~\ref{occupation}. 
The top (bottom) panel corresponds to the top (bottom) panel of Fig.~\ref{occupation}}
\end{center}
\end{figure}

\begin{figure}
\begin{center}
\raisebox{0pt}[2.2cm][0pt]{
\makebox[4cm][l]{\includegraphics[width=4.2cm]{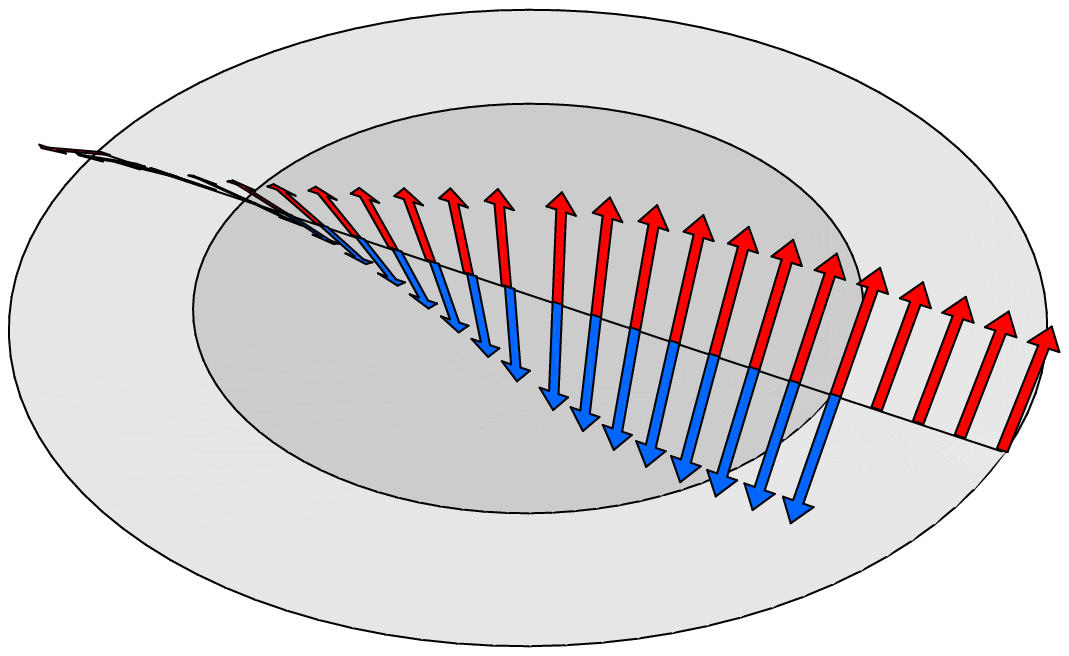}}
\makebox[4cm][r]{\includegraphics[width=4.2cm]{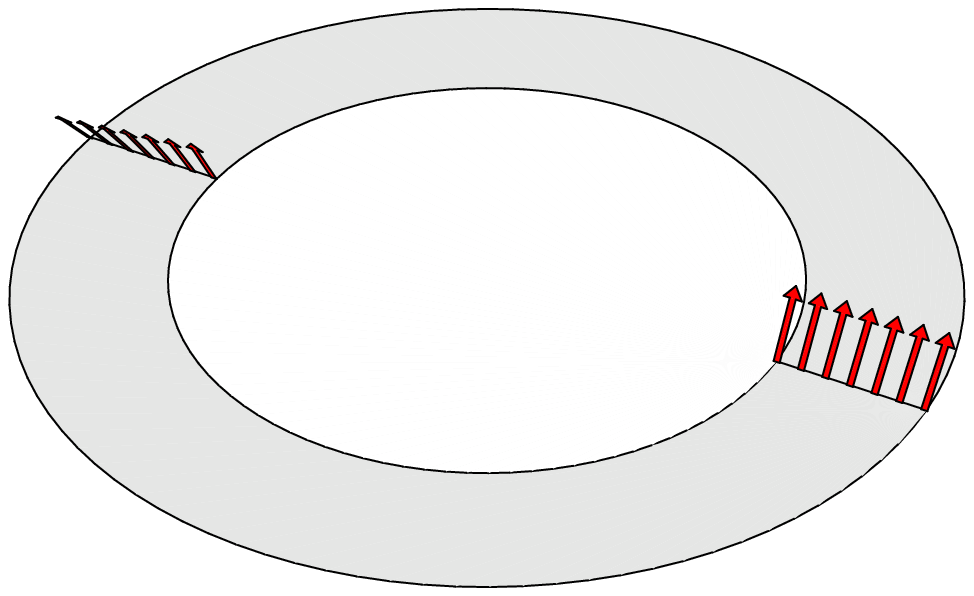}}}
\raisebox{-0.5cm}[1.7cm][0cm]{\includegraphics[width=4.2cm]{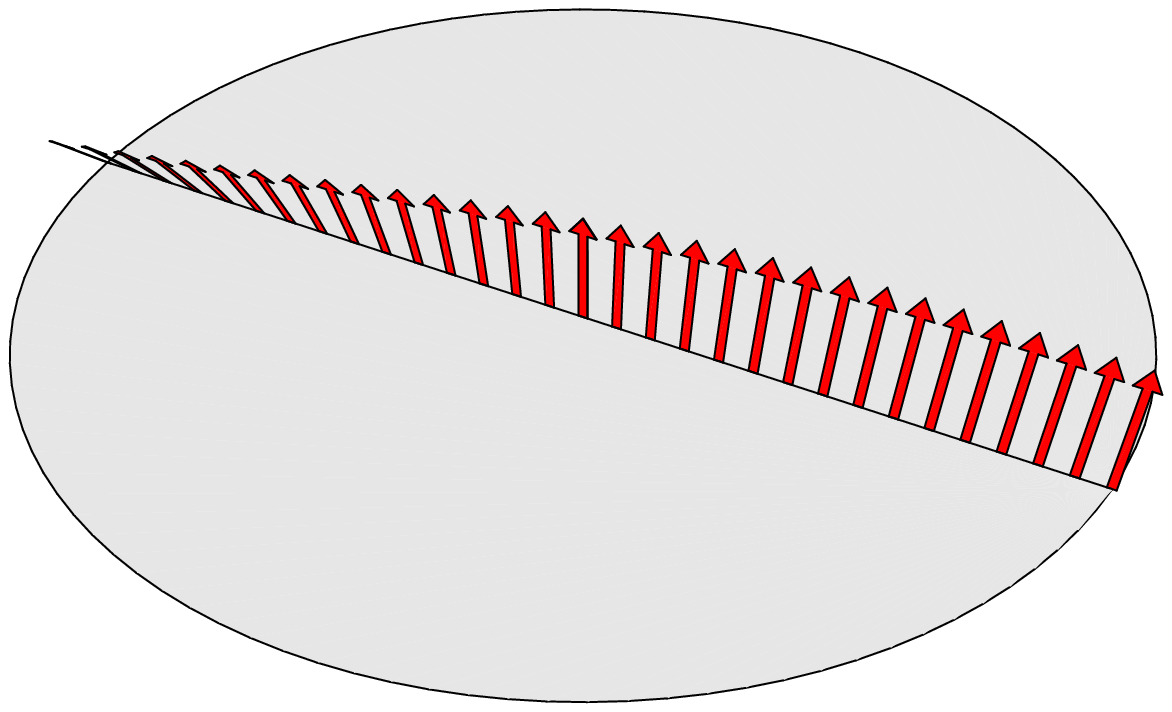}}
\caption{\label{texture_pol} 
Momentum space spin texture for polarized states. 
Top: $\chi=0.4$ and $\chi=2.5$ for $\bar\alpha=0.1$. 
Bottom: Lowest energy polarized state for $\bar\alpha=0.2$ 
(with $\chi=1$ and maximal polarization).}
\end{center}
\end{figure}

At this point the mean field energy can be obtained by minimization with respect 
to both $\chi$ and $\beta_\mathbf{k}$: 
\begin{equation}
\label{HFenergy_min}
{\cal E}_{MF} (\bar{\alpha},r_s) ~= ~ 
\mathrm{min}_{\chi, \beta_\mathbf{k}} 
{\cal E}_\Psi  ~.
\end{equation}
For a fixed $\chi$ minimization with respect to $\beta_\mathbf{k}$ reduces
to finding the lowest lying solutions of the corresponding mean field equation
\begin{eqnarray}
\label{beta_eqn_adim}
\tan{\bar\beta(\y)}=\frac{
\begin{array}{r}
\int_{\scriptscriptstyle{\sqrt{|1-\chi|}}}^{\scriptscriptstyle{\sqrt{1+\chi}}}
\mathrm{d}\y' \int_0^{2\pi} \frac{\y'\, \sin{\bar\beta(\y') \cos{\theta}}}{ \sqrt{{\y'}^2
+ \y^2 - 2 \y \y' \cos{\theta}}}
\mathrm{d}\theta + 4 \pi \bar{\alpha}\,\y\,
\end{array}
}
{\int_{\scriptscriptstyle{\sqrt{|1-\chi|}}}^{\scriptscriptstyle{\sqrt{1+\chi}}}
\mathrm{d}\y' \int_0^{2\pi} \frac{\y'\, \cos{\bar\beta(\y')}}{\sqrt{{\y'}^2+\y^2-2\y \y'
\cos{\theta}}}\,\mathrm{d}\theta} ~,
\end{eqnarray}
here simplified to the relevant case $\gamma_{\bf k}=\phi_{\bf k}+\frac{\pi}{2}$ 
corresponding to a vanishing in plane polarization.
Remarkably this equation is devoid of any explicit dependence on $r_s$, the solution
$\bar \beta$ being only dependent on $\bar \alpha$ and $\chi$.
Notice that $\bar \beta = \frac{\pi}{2}$ is always a solution of Eq.~(\ref{beta_eqn_adim}).
The latter has rather interesting symmetries. For instance one can see that 
$\bar \beta_{1/\chi}(\y)= \bar \beta_{\chi}(\sqrt{\chi}\, \y)$.
Eq.~(\ref{beta_eqn_adim}) makes it manifest how, for this class of solutions, 
the generalized chirality $\chi$ is the central parameter of the problem. 
As an illustration, typical solutions for the universal azimuthal angle 
$\bar\beta(\y)$ are displayed in Fig.~\ref{manybeta}.
There, solutions with $ \bar \beta \neq \frac{\pi}{2}$ correspond to polarized 
states in which the spin polarization points along the $z$-axis. 
Interestingly, these are constructed out of single particle states that are not
solutions of the single particle hamiltonian (\ref{H0}) and display an intriguing 
spin texture in momentum space. 
This is shown in Fig.~\ref{texture_pol} where three possible cases are depicted.

\begin{figure}
\includegraphics[width=0.45\textwidth]{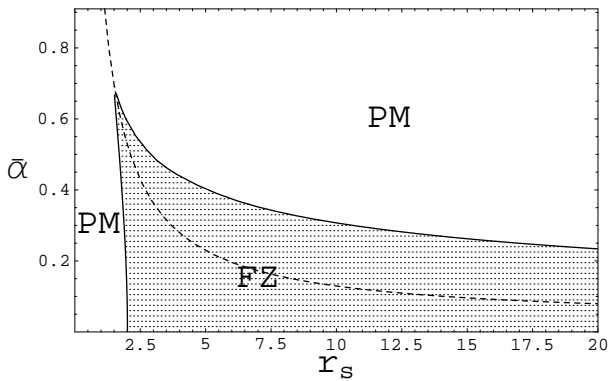}
\caption{\label{phasediagr} Relative phase diagram in the $(r_s,\bar{\alpha})$ 
plane of the isotropically occupied, spatially homogeneous mean field 
solutions. In the shaded region the gas is polarized along the perpendicular direction. 
Along the dashed line the energy of the unpolarized states is minimized by $\chi=1$.}
\end{figure}

\begin{figure}
\begin{center}
\includegraphics[width=0.4\textwidth]{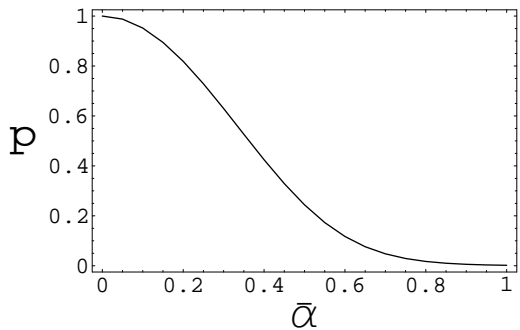}
\caption{\label{spin_pol} Fractional polarization for the transverse 
ferromagnetic solutions (with $\chi=1$), plotted as function of $\bar\alpha$.}
\end{center}
\end{figure}

Further minimization with respect to $\chi$ allows one to compare the 
energies of these homogeneous, symmetrically occupied states. 
This leads to the determination of the boundaries (lines of first order transitions)
between the paramagnetic (PM) and the $z$-axis polarized phases (FZ, dotted area) 
in the relative phase diagram of Fig.~\ref{phasediagr}. Within the ferromagnetic
region the generalized chirality is constant and equals one.
This phenomenon is due to the cuspidal behavior of 
$\mathcal{E}_{x}$ for this value of $\chi$. 
The fractional polarization is readily seen to be given by
\begin{equation}
\label{polarization_eqn}
p\,(\bar\alpha,\chi)=\frac{2 \, \langle \hat S_z \rangle}{\hbar ~ N}=
\int_{\scriptscriptstyle{\sqrt{|1-\chi|}}}^{\scriptscriptstyle{\sqrt{1+\chi}}}\y\,
\cos{\bar\beta(\y)}\,\mathrm{d}\y ~.
\end{equation}
$p$ vanishes for unpolarized states ($\bar \beta = \frac{\pi}{2}$) and acquires
an $r_s$ independent value in the polarized phase for which 
$\bar \beta \neq \frac{\pi}{2}$. 
In general the ferromagnetic polarization is less than one as it can be surmised
from Fig.~\ref{spin_pol}. 
On the other hand $p \rightarrow 1$ for vanishing $\alpha$ as one expects to recover 
the fully polarized Bloch ferromagnetic state of the two dimensional electron liquid in the 
absence of spin-orbit coupling.\cite{TheBook}
A most remarkable feature of this phase diagram is the reentrant paramagnetic 
phase at lower densities, something due to the density dependence of the Rashba 
term (see Eq.~(\ref{Egen})).

We discuss next the possibility of inhomogeneous states. It is clear that in the
low density limit the system will form a Wigner crystal. On the other hand, we 
have found that, as in the absence of spin-orbit coupling, within the mean field 
approximation, homogeneous states are unstable to spin density wave type 
distortions.\cite{AWO} We have been able to construct a rigorous proof of this 
statement valid for all densities. The procedure is patterned after that 
of Ref.~\onlinecite{TheBook}. One assumes the following set of trial single particle 
wave functions:
\begin{equation}
\label{SDWwavefunct} 
\psi_{\bf {k}} ( {\bf r} ) 
\simeq \varphi_{ {\bf k} , +} ( {\bf r} ) + A_{\bf {k} +} 
\varphi_{ {\bf k} + {\bf Q}, +} ( {\bf r} ) +
A_{\bf {k} -} \varphi_{ {\bf k} - {\bf Q}, +} ( {\bf r} ) ~,
\end{equation}
where we have chosen 
${\bf Q} = 2k_F \hat x = \frac{4 \bar \alpha }{a_B } \hat x$ as to connect states
on opposite sides of the Fermi energy and suitable for the case of 
$\epsilon_F=0$ in Fig.~\ref{occupation}.
The corresponding leading change in the total mean field energy is a functional
$\Delta E_{MF}[\delta\rho_{\alpha\beta}]$ of the linear
variations of the matrix elements of the single particle density matrix
$\rho _{\alpha \beta } = \left\langle {\Phi  \left| {\hat a_\alpha ^\dag \hat a_\beta }
\right|\Phi } \right\rangle$ whose expression can be found in Ref.~\onlinecite{TheBook}.
At this point, after some algebra, one finds that the following judicious 
choice of the amplitudes
\begin{equation}
\label{Amplitude} A_{\bf {k} \pm}=
 \left\{
 \begin{array}{lll}
 \frac{\left( {bk_F } \right)^{\frac{3}{2}}}{\ln \frac{2}{b}}
 \frac{| n_{{\bf k} \pm  {\bf Q}} - n_{\bf k} | }{
| { \bf k } \pm  \frac{{\bf Q}}{2} |^{\frac{3}{2}}}
\frac{
\sqrt{| \sin \phi _{ {\bf k} \pm  \frac{{\bf Q}}{2}} |} 
} {
| \cos \phi _{ {\bf k} \pm  \frac{{\bf Q}}{2}} |}& &
bk_F < | {\bf k} \pm  \frac{{\bf Q}}{2} | < ebk_F  \\
 0& & {\rm otherwise,}\\
\end{array} \right.
\end{equation}
where $b$ is an arbitrarily small positive number, leads to a negative 
$\Delta E_{MF}$ even without allowing for momentum space repopulation.

An inspection of our procedure reveals that, within mean field theory, several of 
the effects of the interactions can be described in terms of the concept of 
effective magnetic field, a quantity defined as
\begin{equation}
\label{def_Beff}
\frac{g \mu_B}{2} \vec{B}_{eff} = -\hbar \alpha \, k \, \hat{\phi}_\mathbf{k}
-\frac{1}{2 L^2}\sum_{\mathbf{k}'} ( n_{\mathbf{k}'+}-n_{\mathbf{k}'-} )\,
v_{\mathbf{k}-\mathbf{k}'} \, \hat{s}_{\mathbf{k}'} ~,
\end{equation}
the last term being associated with the exchange energy. As it can be shown, this 
concept can be employed to write the mean field equation (\ref{beta_eqn_adim}) 
at once.

We conclude by noticing that, as in the absence of spin-orbit, one should reasonably
expect correlation effects to change the density range in which the various transitions 
occur. On the other hand the qualitative structure of the phase diagram should remain 
essentially unchanged. We also expect that the interplay of interactions and spin-orbit 
will lead to interesting novel phenomena not only in the transition region to
the Wigner crystal, where the spin polarization of the states will play a major role, 
but also at intermediate densities where interesting in-plane spin structures could be 
stabilized by both exchange and correlations. 
The current analysis only offers a glimpse at the rich physical complexity of this 
problem. 


\begin{thebibliography}{11}
\expandafter\ifx\csname natexlab\endcsname\relax\def\natexlab#1{#1}\fi
\expandafter\ifx\csname bibnamefont\endcsname\relax
  \def\bibnamefont#1{#1}\fi
\expandafter\ifx\csname bibfnamefont\endcsname\relax
  \def\bibfnamefont#1{#1}\fi
\expandafter\ifx\csname citenamefont\endcsname\relax
  \def\citenamefont#1{#1}\fi
\expandafter\ifx\csname url\endcsname\relax
  \def\url#1{\texttt{#1}}\fi
\expandafter\ifx\csname urlprefix\endcsname\relax\def\urlprefix{URL }\fi
\providecommand{\bibinfo}[2]{#2}
\providecommand{\eprint}[2][]{\url{#2}}

\bibitem[{\citenamefont{Tan et~al.}(2005)\citenamefont{Tan, Zhu, Stormer,
  Pfeiffer, Baldwin, and West}}]{tan05}
\bibinfo{author}{\bibfnamefont{Y.-W.} \bibnamefont{Tan}},
  \bibinfo{author}{\bibfnamefont{J.}~\bibnamefont{Zhu}},
  \bibinfo{author}{\bibfnamefont{H.~L.} \bibnamefont{Stormer}},
  \bibinfo{author}{\bibfnamefont{L.~N.} \bibnamefont{Pfeiffer}},
  \bibinfo{author}{\bibfnamefont{K.~W.} \bibnamefont{Baldwin}},
  \bibnamefont{and} \bibinfo{author}{\bibfnamefont{K.~W.} \bibnamefont{West}},
  \bibinfo{journal}{Phys. Rev. Lett.} \textbf{\bibinfo{volume}{94}},
  \bibinfo{pages}{016405} (\bibinfo{year}{2005}).

\bibitem[{\citenamefont{Winkler et~al.}(2005)\citenamefont{Winkler, Tutuc,
  Papadakis, Melinte, Shayegan, Wasserman, and Lyon}}]{winkler05}
\bibinfo{author}{\bibfnamefont{R.}~\bibnamefont{Winkler}},
  \bibinfo{author}{\bibfnamefont{E.}~\bibnamefont{Tutuc}},
  \bibinfo{author}{\bibfnamefont{S.~J.} \bibnamefont{Papadakis}},
  \bibinfo{author}{\bibfnamefont{S.}~\bibnamefont{Melinte}},
  \bibinfo{author}{\bibfnamefont{M.}~\bibnamefont{Shayegan}},
  \bibinfo{author}{\bibfnamefont{D.}~\bibnamefont{Wasserman}},
  \bibnamefont{and} \bibinfo{author}{\bibfnamefont{S.~A.} \bibnamefont{Lyon}},
  \bibinfo{journal}{Phys. Rev. B} \textbf{\bibinfo{volume}{72}},
  \bibinfo{pages}{195321} (\bibinfo{year}{2005}).

\bibitem[{\citenamefont{Asgari et~al.}(2005)\citenamefont{Asgari, Davoudi,
  Polini, Giuliani, Tosi, and Vignale}}]{asgari05}
\bibinfo{author}{\bibfnamefont{R.}~\bibnamefont{Asgari}},
  \bibinfo{author}{\bibfnamefont{B.}~\bibnamefont{Davoudi}},
  \bibinfo{author}{\bibfnamefont{M.}~\bibnamefont{Polini}},
  \bibinfo{author}{\bibfnamefont{G.~F.} \bibnamefont{Giuliani}},
  \bibinfo{author}{\bibfnamefont{M.~P.} \bibnamefont{Tosi}}, \bibnamefont{and}
  \bibinfo{author}{\bibfnamefont{G.}~\bibnamefont{Vignale}},
  \bibinfo{journal}{Phys. Rev. B} \textbf{\bibinfo{volume}{71}},
  \bibinfo{pages}{045323} (\bibinfo{year}{2005}).

\bibitem[{EP2()}]{EP2DS15_05_proc}
\bibinfo{note}{Proceedings of the 16th International Conference on Electronic
  Properties of Two-Dimensional Systems (EP2DS-15), edited by M. P. Lilly, W.
  Pan, R.-R. Du and J. A. Simmons (Elsevier, Amsterdam, 2006).}

\bibitem[{\citenamefont{Giuliani and Vignale}(2005)}]{TheBook}
\bibinfo{author}{\bibfnamefont{G.~F.} \bibnamefont{Giuliani}} \bibnamefont{and}
  \bibinfo{author}{\bibfnamefont{G.}~\bibnamefont{Vignale}},
  \emph{\bibinfo{title}{Quantum Theory of the Electron Liquid}}
  (\bibinfo{publisher}{Cambridge University Press},
  \bibinfo{address}{Cambridge}, \bibinfo{year}{2005}).

\bibitem[{com({\natexlab{a}})}]{comment_generalized_SO}
\bibinfo{note}{An analysis of the properties of the exchange energy in a
  generalized spin-orbit model that is relevant in a number of experimental
  situations can be found in S. Chesi and G. F. Giuliani, cond-mat/0701355
  (2007).}

\bibitem[{ras()}]{rashbaSO}
\bibinfo{note}{Y. A. Bychkov and E. I. Rashba, JETP Lett. {\bf 39}, 78 (1984);
  J. Phys. C {\bf 17}, 6039 (1984).}

\bibitem[{com({\natexlab{b}})}]{comment_dresselhaus}
\bibinfo{note}{The same analysis applies in the presence of (although not in
  concomitance with) a term of the Dresselhaus type. See G. Dresselhaus, Phys.
  Rev. {\bf 100}, 580 (1955).}

\bibitem[{com({\natexlab{c}})}]{comment_previous_work_QP}
\bibinfo{note}{Some early work on the quasiparticle properties of this system
  can be found in G.-H. Chen and M. E. Raikh, Phys. Rev. B {\bf 60}, 4826
  (1999); for a more recent analysis see D. S. Saraga and D. Loss, Phys. Rev. B
  {\bf 72}, 195319 (2005).}

\bibitem[{\citenamefont{Chesi and Giuliani}()}]{unpub_ChesiGFG}
\bibinfo{author}{\bibfnamefont{S.}~\bibnamefont{Chesi}} \bibnamefont{and}
  \bibinfo{author}{\bibfnamefont{G.~F.} \bibnamefont{Giuliani}},
  \bibinfo{note}{unpublished.}

\bibitem[{AWO()}]{AWO}
\bibinfo{note}{A. W. Overhauser, Phys. Rev. Lett. {\bf 4}, 462 (1960); Phys.
  Rev. {\bf 128}, 1437 (1962).}

\end{thebibliography}
\end{document}